\documentclass[11pt,a4paper]{article}

\usepackage[a4paper, margin=2.3cm]{geometry}

\usepackage{url}
\usepackage{listings}
\usepackage{booktabs}
\usepackage{tabularx}

\usepackage[most]{tcolorbox}
\usepackage{enumitem}
\newtcolorbox{contextbox}{
  colback=white,
  colframe=black,
  boxrule=0.6pt,
  arc=0pt,
  left=8pt,
  right=8pt,
  top=8pt,
  bottom=8pt
}

\usepackage{caption}
\usepackage{xcolor}
\newcommand{\graybox}[1]{%
\noindent
\colorbox{yellow!15}{
\parbox{\dimexpr\linewidth-2\fboxsep}{#1}%
}
}

\newcommand{\agent}[1]{\textsf{#1}}
\newcommand{\ark}{\textsf{Ark}}
\newcommand{\arkbench}{\textsf{ArkBench}}

\usepackage{soul}
\sethlcolor{yellow!18}

\usepackage{pgfplots}
\pgfplotsset{compat=1.18}

\usepackage{graphicx} 

\title{Understanding the Architecture of Coding Agents: An Exploratory Study Using a Research Prototype}
\author{Marco Tulio Valente}
\date{Department of Computer Science \\ Federal University of Minas Gerais, Brazil \\
mtov@dcc.ufmg.br}

\begin{document}

\maketitle

\begin{abstract}
\noindent Coding agents have rapidly emerged as the primary interface for AI-assisted software development. However, despite their growing adoption, relatively little is known about their internal architecture, and no systematic architectural description comparable to those available for compilers or operating systems currently exists. This paper addresses this gap by documenting the main architectural components of coding agents, explaining their responsibilities, interactions, and execution flow. To support this effort, we also present \ark\ (Agent Research Kit), a minimal open-source coding agent designed for research and education that preserves the essential architectural mechanisms of modern coding agents while emphasizing simplicity and clarity. We also introduce \arkbench, a lightweight benchmark comprising ten representative software maintenance and evolution tasks. Using {\tt gpt-5.4-mini}, \ark\ successfully solved 8 of the 10 tasks while requiring modest token consumption. Finally, we compare the architecture of \ark\ with those of state-of-the-art coding agents using a recently proposed architectural taxonomy. We hope that both \ark\ and \arkbench\ provide a practical foundation for teaching, research, and experimentation on coding agents.\\[-0.25cm]

\end{abstract}

\section{Introduction}

Coding agents have emerged as one of the most important applications of Large Language Models (LLMs). Unlike first-generation AI programming assistants, which primarily answer questions or generate code snippets, coding agents autonomously carry out complete software development tasks through iterative interaction with an LLM and the local development environment. To this end, they can inspect source code, search for relevant information, run tests, edit files, and progressively refine their solutions until the requested task is completed. Consequently, coding agents are being rapidly integrated into the workflows of developers and organizations, becoming a core component of modern software engineering~\cite{robbes2026agentic}.

Despite their rapid adoption, relatively little is known about the internal architecture of coding agents~\cite{Rombaut2026}. Commercial systems such as Claude Code and Cursor offer sophisticated capabilities but do not disclose their implementations, which makes it difficult to understand their internal design, reuse their architectural solutions, or adopt them as research platforms. Open-source alternatives, including Codex, OpenCode, and Aider, do make their source code available; however, these projects have already grown into large, complex systems comprising numerous modules and are designed primarily for production use rather than architectural clarity. Their documentation likewise focuses mainly on installation and usage, offering limited insight into the architectural principles underlying their design or the responsibilities of their core components. As a result, a systematic and clear architectural description of a modern coding agent is still lacking.

This paper addresses this gap by \hl{documenting and explaining the main architectural components of coding agents}. Specifically, we describe the responsibilities of these components, how they interact, and the execution flow that enables an agent to autonomously solve programming tasks. Our motivation stems from the rapid emergence of coding agents as the primary interface for software development, much as compilers became indispensable with the advent of high-level programming languages in the 1970s. However, while compiler architectures have been thoroughly documented for decades~\cite{aho2006, appel2002}, no comparable architectural reference exists for coding agents. Another source of inspiration, from a different domain, was the Minix operating system, which was designed to make operating system architecture more accessible to students and researchers~\cite{tanenbaum1987}.

Furthermore, \hl{to ground our study in current practice, we implemented {\sf Ark} (Agent Research Kit)}, a minimal, open-source coding agent designed for research and education. Rather than maximizing functionality, \ark\ emphasizes architectural clarity while preserving the essential mechanisms of modern coding agents, including the agentic loop, structured interaction with LLMs, tool invocation, agent memory, workspace management, and the safe application of code modifications. To evaluate \ark's capabilities, we also developed \arkbench, a lightweight benchmark comprising ten representative software maintenance and evolution tasks, on which \ark\ successfully solved 8 of the 10 tasks. Finally, we compare the architecture of \ark\ with those of state-of-the-art coding agents using a recently proposed taxonomy for coding agent architectures~\cite{Rombaut2026}, highlighting the architectural design choices that distinguish \ark\ from existing systems.

The main contributions of this paper are fourfold: (1) a systematic description of the main architectural components of coding agents and the interactions among them; (2) \ark, a fully documented, open-source minimal coding agent designed as a platform for education, research, and experimentation; (3) \arkbench, a lightweight benchmark comprising representative software maintenance and evolution tasks; and (4) a practical foundation for future research on the design, implementation, and evaluation of coding agents.

The remainder of this paper is organized as follows. Section~\ref{sec:methodology} describes the methodology adopted to design and implement \ark. Section~\ref{sec:architecture} presents the main architectural components of coding agents and explains how they are realized in \ark. Section~\ref{sec:evaluation} introduces \arkbench\ and reports the experimental evaluation of \ark. Section~\ref{sec:comparison} compares the architecture of \ark\ with state-of-the-art coding agents using a recently proposed taxonomy. Section~\ref{sec:related-work} discusses related work. Finally, Section~\ref{sec:conclusion} concludes the paper and outlines directions for future research.

\section{Methodology}
\label{sec:methodology}

To investigate the architecture of a coding agent, we implemented our own agent, called \ark. In the initial commits of this implementation, our goal was simply to follow the ReAct pattern proposed by Yao et al.~\cite{react-iclr}. During this early stage, we also relied on technical documentation published by leading AI companies, particularly an article by OpenAI describing the internal operation of Codex~\cite{openai2026codexloop} and a similar article by Anthropic describing the architecture of Claude Code~\cite{anthropic2026agentsdk}.

After studying the aforementioned references, the implementation was carried out iteratively by the author. Rather than developing all components at once, new functionality was introduced incrementally and validated through small-scale usage scenarios involving typical software engineering tasks, such as bug fixing, refactoring, and feature implementation. This iterative process made it possible to evaluate and refine the agent's main architectural components, including the agentic loop, the interaction protocol with the LLM, agent memory, context management, and the available tools. During the early stages of development, two open-weight LLMs from the Qwen family ({\tt qwen2.5-coder:14b} and {\tt qwen3-coder:30b}) were used locally. However, their performance was generally unsatisfactory, particularly on tasks requiring strict adherence to the agent's interaction protocol and reliable patch generation. Consequently, in the final stage of development, these models were replaced with a commercial LLM ({\tt gpt-5.4-mini}), which exhibited more stable behavior and consistently better performance throughout our experiments.

\begin{figure}[ht]
\centering
\begin{tikzpicture}
\begin{axis}[
    xbar,
    width=\linewidth,
    height=6cm,
    xmin=0,
    xlabel={Number of Commits},
    symbolic y coords={
        Refactor,
        Docs,
        Feature,
        Bug Fix,
        Tests
    },
    ytick=data,
    y dir=reverse,          % maior no topo
    bar width=10pt,          % barras mais grossas
    y=0.65cm,               % diminui espaçamento entre barras
]
\addplot coordinates {
    (55,Refactor)
    (48,Docs)
    (19,Feature)
    (11,Bug Fix)
    (7,Tests)
};
\end{axis}
\end{tikzpicture}
\caption{Number of commits used in the implementation of \ark, by task category.}
\label{fig:commits-ark}
\end{figure}
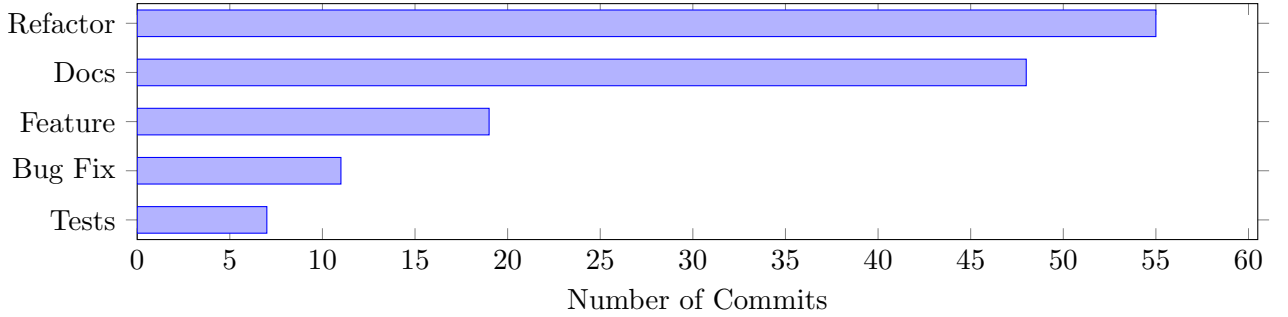

The development of \ark\ was itself supported by the Codex coding agent. In other words, we used one coding agent to implement almost the entire codebase of another. This decision provided valuable insights into the practical operation of coding agents. Nevertheless, all code generated by Codex was carefully reviewed by the author before being incorporated into the project.
By the end of development, the project comprised 150 commits distributed across several categories. As shown in Figure~\ref{fig:commits-ark}, 55 commits (37\%) were devoted to refactoring the codebase, improving its modularity and readability. Another 48 commits (32\%) focused on maintaining comprehensive and up-to-date documentation, particularly the {\tt README.md} file. These results show that, as the architecture evolved, substantial effort was invested in simplifying the implementation and improving its documentation, consistent with the project's goal of serving as a platform for research and education. The remaining commits correspond primarily to feature implementation (13\%), bug fixes (7\%), test development (5\%), and configuration tasks (3\%). 
In summary, Figure~\ref{fig:commits-ark} shows that, \hl{although the code was generated by an LLM, considerable engineering effort, particularly through refactoring, was required to ensure high code quality}. This was expected given the exploratory nature of the development process, in which the architecture gradually emerged through multiple rounds of trial and error~\cite{softengbook}.

Table~\ref{tab:arquivos-ark} provides an overview of the source files that comprise \ark. As the table shows, the implementation meets its primary design goal of remaining compact and easy to understand. The entire codebase consists of just 13 Python source files totaling 1,471 lines of code (LOC). Furthermore, the project relies on only two external dependencies: {\tt openai}, for accessing LLM APIs, and {\tt pytest}, for running the test suite.

\begin{table*}[ht]
\centering
\small
\caption{Main source files of \ark}
\label{tab:arquivos-ark}
\begin{tabular}{lrp{11cm}}
\hline
\textbf{File} & \textbf{LOC} & \textbf{Description} \\
\hline
{\tt agentic\_loop.py}   & 237 & Implements the main loop, tool calls, memory updates, and stopping conditions. \\
{\tt inputs.py}          & 181 & Loads prompts/configuration, selects the task, and prepares the workspace. \\
{\tt models.py}          & 179 & Interfaces with the LLM, handling requests and responses. \\
{\tt tools.py}           & 163 & Defines tools for file inspection, modification, and testing. \\
{\tt patches.py}         & 346 & Repairs, validates, previews, and applies patches. \\
{\tt protocol.py}        & 70  & Defines the LLM response format and validates actions and arguments. \\
{\tt finish\_handler.py} & 67  & Handles the {\tt finish} action and final patch checks. \\
{\tt traces.py}          & 82  & Records traces, tool usage, token counts, and run summaries. \\
\hline
\end{tabular}
\end{table*}

The codebase also includes a test suite comprising 50 unit tests distributed across eight test files. The tests were generated automatically with the assistance of Codex during development and subsequently incorporated into the project's validation infrastructure. In the current version, all tests pass successfully, and line coverage over the Python modules in {\tt src/ark} reaches 100\%.\\

\graybox{
{\em Availability:} \ark\ is publicly available, under a MIT license, at \url{https://github.com/mtov/ark}.}.

\section{Architecture of Coding Agents}
\label{sec:architecture}

\begin{figure}[ht]
\centering
\includegraphics[width=0.8\columnwidth]{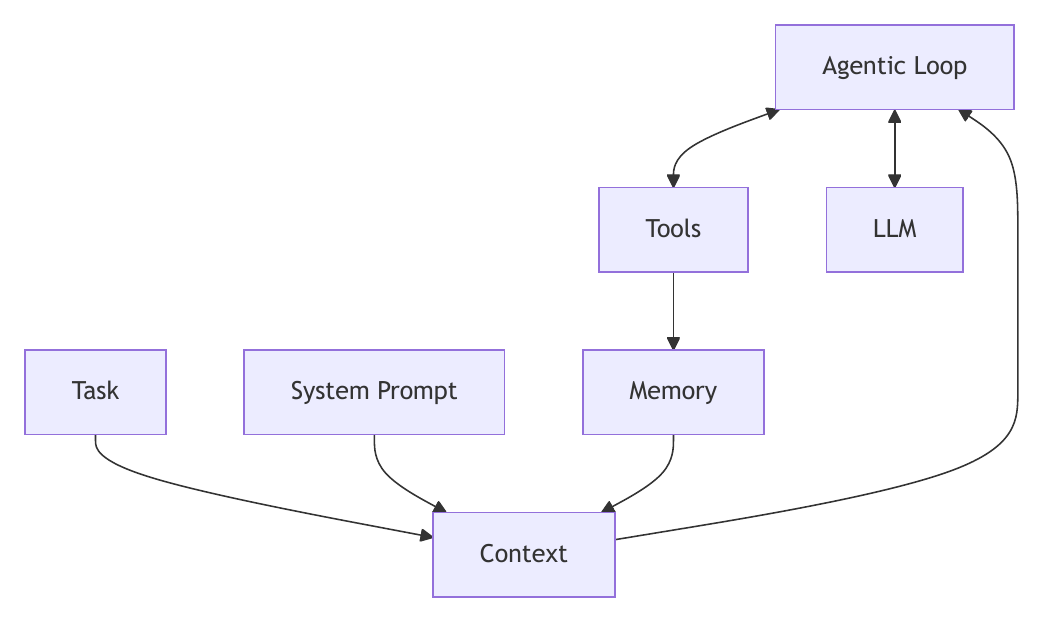}
\caption{Architecture of a Coding Agent}
\label{fig:architecture}
\end{figure}

Figure~\ref{fig:architecture} illustrates the main architectural components of a coding agent and the relationships among them. As shown in the figure, the {\bf agentic loop} plays a central role, coordinating the interaction between the {\bf LLM} and the other components. Before each interaction with the LLM, the agent constructs a {\bf context} consisting of the {\bf system prompt}, the task description, and the information stored in the agent's {\bf memory}. The LLM may then request the execution of a {\bf tool}, whose output is recorded in the agent's {\bf memory} and incorporated into the context for the next iteration. This process repeats until the LLM determines that the task is complete and produces the final response. The following subsections examine each of these components in detail. For clarity, we first present the architectural role of each component and then describe how it is implemented in our reference implementation.

\subsection{Agentic Loop}
\label{sec:agentic-loop}

\begin{figure}[ht]
\centering
\includegraphics[width=0.6\linewidth]{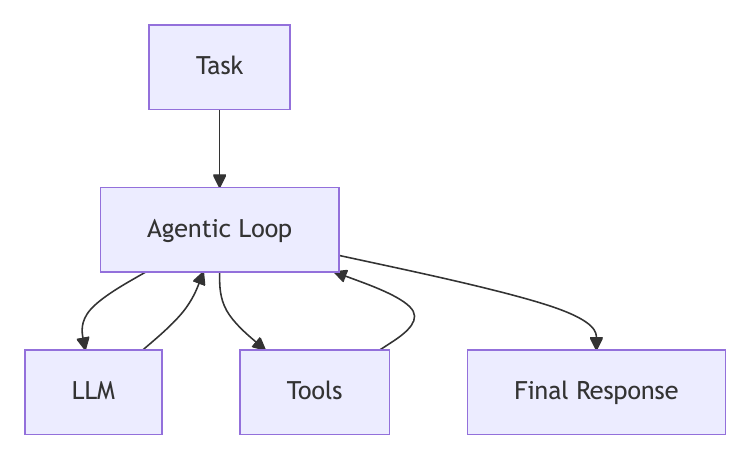}
\caption{Agentic Loop}
\label{fig:loop}
\end{figure}

The core component of a coding agent is the \textbf{agentic loop}. Through this loop, the agent repeatedly interacts with the LLM to solve a programming task. Rather than sending a single prompt and waiting for a final response, the agent and the LLM engage in multiple rounds of interaction until the requested task has been completed, as illustrated in Figure~\ref{fig:loop}. 

Another important aspect is that coding agents typically follow the ReAct (\textit{Reason + Act}) interaction paradigm~\cite{react-iclr,huyen2025aiengineering}. Under this paradigm, the LLM alternates between reasoning and requesting the execution of actions. More specifically, the {\bf system prompt} instructs the LLM to behave as follows:

\begin{itemize}

\item The LLM must first reason about the problem. To do so, it decomposes the original task into smaller subtasks and addresses them one at a time.

\item The LLM may also request the execution of local programs, referred to as \textbf{tools}. Typical examples include listing, reading, searching, and editing files, running tests, executing Git commands, and applying patches.

\end{itemize}

After executing a tool, the agent returns its output to the LLM. Based on this feedback, the LLM decides which subtask to perform next, which may involve requesting the execution of another tool. This process continues until the LLM determines that the original task has been completed successfully. At that point, the agentic loop terminates, and the LLM's final response is returned to the agent and presented to the user.

Because the agentic loop is the core component of a coding agent, Listing~\ref{lst:loop} presents its complete implementation in \ark. Despite its central role, the loop consists of only 18 lines of code. Execution begins by creating the agent's {\bf memory} (line 2), which records the tools executed during each iteration together with their outputs. The agent then enters a loop (line 4), bounded by a maximum number of iterations. In each iteration, the function \texttt{get\_next\_tool\_request} queries the LLM to determine the next tool to execute (line 5). If the requested action is \texttt{finish}, the agent returns the final response and terminates (lines 7--11). Otherwise, the requested tool is executed locally (line 15), and its output is recorded in the agent's memory (line 16). If the iteration limit is reached, the agent terminates with an error (line 18).

\begin{lstlisting}[caption={Agentic Loop},label={lst:loop}]
def agentic_loop(config: AgentConfig) -> LoopResult:
    memory = Memory()

    for iteration in range(1, MAX_ITERATIONS + 1):
        tool_request = get_next_tool_request(config, memory)

        if tool_request.name == "finish":
            output = handle_finish(config, memory, iteration, tool_request)
            if output is None:
                continue
            return LoopResult.success(output)

        print_iteration_action(iteration, tool_request)

        tool_result = run_tool(tool_request, config)
        memory.append(iteration, tool_request, tool_result)

    return LoopResult.max_iterations_reached()
\end{lstlisting}

\subsection{System Prompt}
\label{sec:system-prompt}

LLMs typically receive two types of prompts: a \textbf{system prompt} and a \textbf{user prompt}. The system prompt provides high-level instructions that define the model's behavior, including its role, constraints, and response format. The user prompt, by contrast, contains the specific request to be addressed, such as a question, a programming task, or a command. These prompts are passed as parameters in API calls to the LLM, as illustrated below:\\

\begin{lstlisting}
messages = [
    {"role": "system", "content": "You are a coding assistant."},
    {"role": "user",   "content": "Fix the bug in add()."}
]
\end{lstlisting}

The {\bf system prompt} is therefore a fundamental design decision in a coding agent, as it defines the behavior the LLM must follow throughout its interaction with the agent. In particular, the system prompt should: (1) instruct the LLM to behave as a coding agent; (2) specify the interaction protocol, including the required response format; (3) describe the tools the LLM is allowed to invoke; (4) define operational constraints, such as permissions and file access; (5) guide the LLM's decision-making process, for example, by specifying when it should read files, run tests, or conclude a task; and (6) define the termination criteria that determine when the LLM should produce the final response. Because of its central role, Listing~\ref{lst:system-prompt} presents the complete system prompt used by \ark, which is stored in the file {\tt config/system\_prompt.txt}. As shown, the system prompt is longer than the agentic loop itself, comprising 33 lines. It begins by defining the LLM's role as a programming assistant (line 1) and specifying a ReAct-style interaction protocol, including the required response format and the actions---that is, tools---the LLM is allowed to invoke (lines 2--9). Next, it establishes rules governing tool usage and workspace access, restricting operations to the available files and discouraging redundant actions (lines 11--13). The prompt also imposes constraints to ensure consistent execution, such as requiring a file to be read before it can be modified and allowing the task to be completed only when sufficient evidence has been gathered (lines 15--23). Finally, it requires the final output as a diff patch.

\begin{lstlisting}[language={},   breaklines=true,
    breakindent=0pt,basicstyle=\ttfamily\footnotesize,
    caption={System Prompt},label={lst:system-prompt}
]
You are a coding assistant.
Respond with exactly:
Thought: your brief reasoning
Action: list_files, read_file, find_text, run_tests, or finish
Action Input: the file path, the search input, blank for run_tests, or a unified diff patch for finish

Choose exactly one action.
Stop after that action.
Never include Observation.

Treat the workspace root provided in the prompt as the only valid root for tool paths.
Use only files and directories that were returned by tools or already appear in agent history.
Do not invent paths or guessed roots such as `/`, `/workspace`, or `/config`.

Do not repeat the same read or search unless you still need missing information.
If you already read a file or ran the same search, use the earlier result from agent history and continue from it instead of repeating it, unless you need a different file or query.
Prefer files and searches already listed in agent history.
Before returning a patch, make sure you have read every file you are modifying.
Do not edit, rename, or update a file that you have not read in the current run.

Do not use Action: finish before you have enough information from the workspace to complete the task.
When you use Action: finish, return only a unified diff patch in Action Input.
Do not claim that the patch was applied.
Use real unified diff hunk headers with line ranges, for example:
--- a/file.py
+++ b/file.py
@@ -1 +1 @@
-old_name
+new_name

Use `find_text` with this format: search text | relative/or-known/workspace/path
Use `run_tests` when the task asks you to validate behavior, fix behavior, or confirm a change.
If workspace instructions are provided in the user prompt, follow them as local guidance for that workspace.
\end{lstlisting}

Designing the system prompt of a coding agent is largely an empirical process. For example, the first version of \ark's system prompt consisted of only 13 lines and simply instructed the LLM to follow the ReAct paradigm while listing the tools it was allowed to invoke. As the agent evolved, the prompt was progressively refined. At one point, for instance, we observed that the LLM occasionally generated patches that modified files it had never read. This behavior increased the likelihood of inconsistent patches and application failures. To address this issue, the prompt was revised to explicitly require the LLM to read every file before proposing modifications to it (line 18 of Listing~\ref{lst:system-prompt}).

Another important aspect is that the system prompt is sent to the LLM with every request issued by the agent. Consequently, its length directly affects token consumption and, therefore, execution cost and latency. For this reason, the system prompt should be kept as concise as possible without sacrificing the instructions essential for the correct operation of the agent.

\subsection{Tools}
\label{sec:tools}

One of the primary responsibilities of a coding agent is to provide the LLM with the information needed to complete the user's requested task. To obtain this information, the agent executes local tools at the LLM's request and returns their output to the model, as discussed previously. The tools that an LLM is allowed to invoke must be listed in the system prompt, as illustrated in Listing~\ref{lst:system-prompt} (lines 4--5 and 31--32). In the current implementation of \ark, the following tools are available:

\begin{itemize}

\item {\tt list\_files}: Lists the files in a specified directory, allowing the agent to explore the project's directory structure.

\item {\tt read\_file}: Returns the contents of a specified file, enabling the agent to inspect the project's source code.

\item {\tt find\_text}: Searches for a given string within the files of a specified directory. It returns all matching occurrences, along with their file paths and line numbers, allowing the agent to locate code relevant to a particular task.

\item {\tt run\_tests}: Executes the project's test suite and returns its output, enabling the agent to verify whether the proposed modifications preserve the system's expected behavior.

\item {\tt finish}: Terminates the agent's execution and returns a patch containing the  modifications proposed by the LLM to implement the user's requested task.

\end{itemize}

In \ark, these tools are implemented directly in Python, which simplifies their integration with the other system components and provides greater control over the agent's behavior. In other coding agents, however, tools may also wrap external programs, such as operating system utilities or version control commands like {\tt git}.

\subsection{Memory}
\label{sec:memory}

A coding agent maintains a data structure, referred to as {\bf memory}, that records the history of its execution and its interactions with the LLM. Depending on the implementation, this memory may include the task description, the responses generated by the LLM, the tools that have been executed, their arguments, and the corresponding outputs. In \ark, the memory primarily records the tools executed by the agent, the arguments passed to each tool, and the resulting outputs. As discussed next, this information is used to construct the context sent to the LLM in each new request and to prevent the agent from repeating actions that have already been performed.

\begin{figure*}[ht]
\centering
\includegraphics[width=0.9\linewidth]{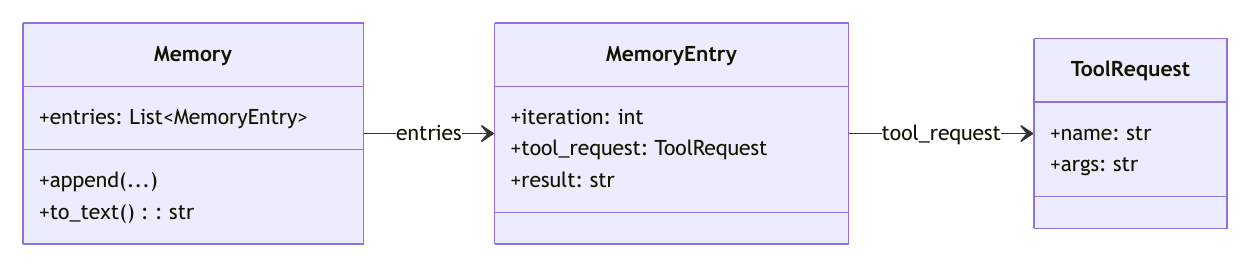}
\caption{Classes implementing an \ark's memory.}
\label{fig:memory}
\end{figure*}

Figure~\ref{fig:memory} illustrates the classes used to implement \ark's memory. The {\tt Memory} class maintains a list of {\tt MemoryEntry} objects, each representing one iteration of the agentic loop. Each entry records the iteration number, the executed tool (represented by a {\tt ToolRequest} object), and the output produced by that tool. Depending on the tool, this output may consist of the contents of a file, the results of a text search, or the output generated by executing the project's test suite.

In the current implementation of \ark, the agent's memory is maintained entirely in main memory and is not persisted across executions. Consequently, each execution starts with an empty memory. This design choice simplifies the implementation and is well suited to scenarios in which executions are independent. However, it also prevents the agent from reusing information gathered during previous executions. Thus, a natural extension is to persist the agent's memory to disk at the end of each execution, allowing it to be restored in future executions and introducing the concept of {\bf sessions}. This mechanism would enable the agent to resume interrupted tasks and reuse knowledge acquired during previous executions.

\subsection{Context}
\label{sec:context}

LLMs are {\em stateless} systems; that is, they do not retain information from previous interactions with a coding agent. Consequently, before each request, the agent must reconstruct and resend the information most relevant to the current execution, such as the tools that have been invoked and their corresponding outputs. Together, this information forms the {\bf context} provided to the LLM, enabling it to determine the next step toward completing the task.

In addition to the system prompt, every request sent by \ark\ to the LLM includes the following contextual information (see also Figure~\ref{fig:context}):

\begin{enumerate}
\item The description of the task to be performed, provided by the user in a {\tt prompt.txt} file located in the project's root directory.

\item Workspace-specific instructions, as defined in a file called {\tt AGENTS.md}, located in the project's root directory. When present, the contents of this file are incorporated into the context, allowing the agent's behavior to be specialized for the particular project without modifying the system prompt. For example, an {\tt AGENTS.md} file may describe project conventions, build and test commands, and other project-specific development guidelines~\cite{agent2026-claude-code,santos2026agentsmd}.

\begin{figure}[ht]
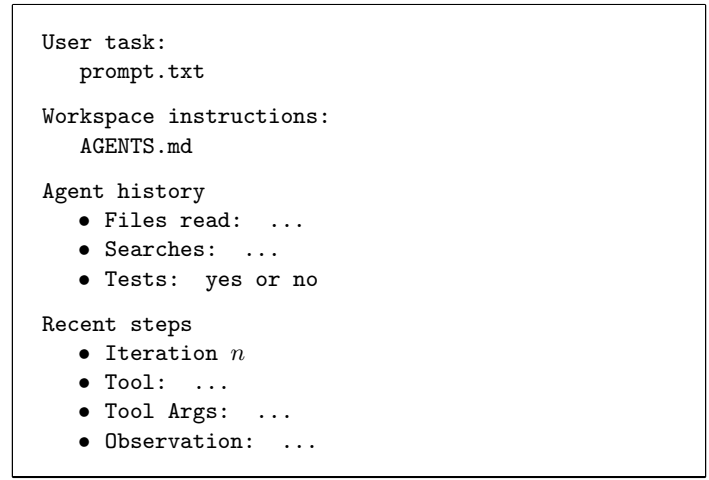

\input figs/fig-contexto.tex
\caption{Context information}
\label{fig:context}
\end{figure}

\item An automatically generated summary of the agent's current state, including the files that have already been read, the text searches that have been performed, and whether the test suite has been executed. This summary enables the LLM to quickly recover the overall execution state without reconstructing it from the interaction history described next.

\item A history of the agent's most recent interactions, derived from its memory, including the tools that were executed, their arguments, and the corresponding outputs. Because LLMs have a limited context window, resending the entire execution history at every iteration is impractical. Instead, \ark\ includes only the four most recent iterations. This design decision reduces the number of tokens sent to the LLM, thereby lowering both cost and latency while still providing the information necessary for successful task completion.

\end{enumerate}

\subsection{Error Handling}

The primary errors that a coding agent must handle arise during its interaction with the LLM. In \ark, two main classes of errors are addressed: inconsistencies in messages that are expected to follow the ReAct protocol and problems in the patches generated by the LLM. The following subsections describe how each of these error-handling mechanisms is implemented.

\subsubsection{ReAct Protocol Errors}
\label{sec:errors}

The \ark\ system prompt defines the fields that must appear in every response returned by the LLM (lines 2--9 of Listing~\ref{lst:system-prompt}). Specifically, responses are required to follow the ReAct format and include three fields: {\it Thought}, {\it Action}, and {\it Action Input}. The {\it Thought} field briefly records the reasoning behind the requested action. The {\it Action} field specifies the single tool to be executed by the agent. Finally, {\it Action Input} provides the arguments required to execute that tool, such as a file path, a search string, or a patch. The system prompt also explicitly states that responses must not include an {\it Observation} field. In ReAct-based agents, an observation corresponds to the output produced by the tool after it has been executed by the agent. In practice, however, LLMs do not always follow this protocol. For example, the model may omit the {\it Action} field, incorrectly include an {\it Observation} field in its response, or produce a free-form answer that does not conform to the required structure.

Accordingly, \ark\ validates every response generated by the LLM. Whenever a response does not conform to the ReAct protocol, the agent returns the following prompt asking the LLM to correct its previous response:\\

\begin{lstlisting}
Your previous response was invalid. 
  Respond using only:
  Thought: ...
  Action: ...
  Action Input: ...
  Do not include Observation.
\end{lstlisting}

If the repair attempt also fails, \ark\ terminates execution and reports the error to the user. In the current implementation, only a single retry is performed.

\subsubsection{Patch Errors}
\label{sec:patch-errors}

During the development of \ark, we observed that the final patch generated by the LLM was not always syntactically valid, even when the intended code changes were correct. For this reason, \ark\ validates the patch before applying it. During this validation, the agent also automatically repairs minor inconsistencies that would otherwise prevent the patch from being applied successfully.

A common example occurs when the LLM generates the correct patch content but reports incorrect line counts in the header of a hunk (a {\it hunk} is a contiguous block of modified lines within a patch). Suppose, for example, that the change simply replaces the line {\tt return 1} with {\tt return 2}, together with three lines of surrounding context. If the LLM generates the hunk header as {\tt @@ -18,4 +18,4 @@} instead of {\tt @@ -18,3 +18,3 @@}, the patch will be rejected, even though the intended modification is correct. In such cases, \ark\ automatically recomputes the line counts from the hunk contents and reconstructs the patch so that it can be applied successfully. Patch validation is performed using the \texttt{git apply --check} command, whereas patch application is carried out with \texttt{git apply}.

\subsection{Security Mechanisms}
\label{sec:security}

As discussed in Section~\ref{sec:tools}, coding agents may execute programs on the user's local machine. Consequently, they must implement security mechanisms that restrict these operations and reduce the risk of unintended or potentially harmful modifications. In the current implementation of \ark, the following security mechanisms are provided:

\begin{itemize}

\item { Temporary workspace isolation.} \ark\ never modifies the original project directory provided by the user. Before execution begins, it creates a copy of the project in a temporary directory named {\tt ark-workspace}. All file reads, test executions, patch validation, and code modifications are performed exclusively within this temporary workspace.

\item {Path restrictions.} All file and directory access tools validate the requested paths to ensure that they remain within the temporary workspace. As a result, the agent cannot use these tools to read or modify arbitrary files on the host machine.

\item {Restricted tool set.} \ark\ exposes only a predefined set of tools to the LLM. Consequently, the model cannot request arbitrary operations on the local machine.

\item {Fixed test command.} The {\tt run\_tests} tool does not allow the LLM to choose which testing framework or command to execute. Instead, \ark\ always runs the fixed command {\tt pytest}, thereby reducing the risks associated with arbitrary command execution on the local machine.

\item {Patch validation and user approval.} Before any modification is applied to the workspace, the patch generated by the LLM is validated both syntactically and structurally, as described in Section~\ref{sec:patch-errors}. After successful validation, \ark\ displays both the command that will be executed and the proposed patch, requesting explicit user approval. Consequently, no changes are applied to the workspace without the user's consent. Once approved, the patch is applied using the standard {\tt git apply} command.

\end{itemize}

\subsection{Tracing}
\label{sec:tracing}

Tracing is an important component of coding agents because it provides a record of the interactions between the agent and the LLM, as well as the actions performed while solving a task. Since the agent's behavior emerges from multiple iterations, tool invocations, and intermediate decisions, tracing facilitate debugging, auditing, and failure analysis. In addition, tracing helps identify issues such as malformed LLM responses, protocol repair attempts, commands proposed to the user, and failures during patch application or test execution.

In \ark, tracing is implemented through a plain-text file, {\tt agent\_trace.log}, which records the requests sent to the LLM, their responses, validation errors,  repair attempts, and events related to patch handling, including patch proposals, user approval or rejection, and final application.

\subsection{LLM Integration}
\label{sec:llm-integration}

LLM integration is an essential component of a coding agent, as it is responsible for sending requests to the model and receiving its responses while abstracting the details of the underlying API. In \ark, this functionality is implemented by a dedicated module that loads the model configuration from a {\tt config.json} file, sends {\tt system} and {\tt user} messages to the LLM, receives the model's responses, collects token usage statistics, and records all interactions through the tracing mechanism. The module also allows users to configure the LLM provider, model name, and endpoint URL (provided it is compatible with the OpenAI API), and the environment variable that stores the API key.

\subsection{Summary}

Table~\ref{tab:agent-components} summarizes the main architectural components of coding agents and their respective responsibilities, as described in this section and realized in \ark.

\begin{table*}[ht]
\centering
\small
\caption{Main architectural components of coding agents}
\label{tab:agent-components}
\begin{tabular}{p{3.5cm} p{11.5cm}}
\toprule
\textbf{Component} & \textbf{Description} \\
\midrule

Agentic Loop &
Coordinates LLM and tool calls until task completion. \\

System Prompt &
Defines LLM behavior, constraints, and interaction format. \\

Tools &
Provide operations to inspect, test, and modify software artifacts. \\

Memory &
Stores actions, observations, and outcomes from previous iterations. \\

Context &
Aggregates information passed to the LLM at each iteration. \\

Error Handling &
Handles invalid outputs, malformed patches, and execution failures. \\

Security Mechanisms &
Restrict and validate potentially unsafe operations. \\

Tracing &
Records interactions and runtime events for analysis and debugging. \\

LLM Integration &
Manages model configuration, communication, and responses. \\

\bottomrule
\end{tabular}
\end{table*}

\section{Experimental Evaluation}
\label{sec:evaluation}

This section presents the evaluation of \ark. We first introduce \arkbench, a lightweight benchmark specifically designed to assess coding agents on software maintenance and evolution tasks. We then report the results obtained by \ark\ on this benchmark and discuss its overall performance.
Our primary goal with this benchmark is not to compare \ark\ with state-of-the-art coding agents. Instead, \hl{our focus is to demonstrate that the proposed architecture resulted in a functional prototype} capable of solving tasks that emulate those performed by developers in their daily work.

\subsection{ArkBench: A Lightweight Benchmark for Coding Agents}
\label{sec:ark-bench}

To evaluate \ark, we developed \arkbench, a small benchmark for coding agents focused on software maintenance and evolution. The benchmark was automatically generated by ChatGPT~5.5 from a set of design rules specified by the author, defining the application domain, programming language, project organization, task categories, complexity level, evaluation criteria, and coding style. The resulting source code, prompts, and tests were then manually reviewed to ensure correctness, consistency, and pedagogical quality. Unlike benchmarks based on large open-source projects, \arkbench\ is intentionally compact, inexpensive to execute, and easy to understand. Despite that, it provides realistic maintenance tasks that require an agent to understand an existing code base, locate relevant files, modify the implementation, and validate the resulting changes. The entire benchmark is implemented as clean, procedural Python code, avoiding unnecessary abstractions, external frameworks, databases, network communication, and accidental complexity.

The benchmark is based on a small e-commerce application. The system is organized into files responsible for product management, order processing, inventory, pricing, validation, filtering, formatting, and report generation, while data is represented using native Python structures such as dictionaries and lists. As summarized in Table~\ref{tab:ark-bench-modules}, each file encapsulates a single well-defined responsibility, resulting in a compact and easy-to-understand code base that still requires agents to navigate multiple files during maintenance tasks.

\begin{table}[ht]
\centering
\small
\caption{Main files of the mini e-commerce application used in \arkbench.}
\label{tab:ark-bench-modules}
\begin{tabularx}{\linewidth}{lX}
\toprule
\textbf{File} & \textbf{Responsibility} \\
\midrule
\texttt{products.py} &
Product catalog management, including pagination and product search \\

\texttt{orders.py} &
Order creation and processing \\

\texttt{inventory.py} &
Inventory availability and stock management \\

\texttt{pricing.py} &
Subtotal, discount, and total price calculations \\

\texttt{validators.py} &
Validation of customer, product, and order data \\

\texttt{filtering.py} &
Filtering operations over product collections \\

\texttt{formatting.py} &
Formatting of monetary values and textual output \\

\texttt{reports.py} &
Generation and export of sales reports \\
\bottomrule
\end{tabularx}
\end{table}

As presented in Table~\ref{tab:ark-bench-tasks}, \arkbench, contains ten independent maintenance tasks organized into three categories: three bug fixes, three refactorings, and four feature implementations. Each task is self-contained, includes its own copy of the project, and requires modifications to approximately two to four source files. 

\begin{table}[ht]
\centering
\small
\caption{Tasks included in \arkbench.}
\label{tab:ark-bench-tasks}
\begin{tabularx}{\linewidth}{lX}
\toprule
\textbf{Task} & \textbf{Description} \\
\midrule

\texttt{bugfix-01} & Fix an off-by-one error in catalog pagination \\
\texttt{bugfix-02} & Reject orders without a customer or products \\
\texttt{bugfix-03} & Fix the discount threshold calculation \\

\addlinespace

\texttt{refactor-01} & Extract duplicated subtotal calculation into a shared function \\
\texttt{refactor-02} & Move the price-formatting function to a dedicated file \\
\texttt{refactor-03} & Rename a public product-search function and update its usages \\

\addlinespace
\texttt{feature-01} & Add product filtering by price range \\
\texttt{feature-02} & Add product sorting by price \\
\texttt{feature-03} & Export order reports in CSV format \\
\texttt{feature-04} & Group sales by product category \\

\bottomrule
\end{tabularx}
\end{table}

Each task follows the same directory structure. The file \texttt{prompt.txt} describes the requested maintenance activity in the style of a GitHub issue, without identifying the files to be modified. The \texttt{tests} directory contains the public test suite available to the agent, while the \texttt{evaluation} directory contains hidden post-task tests and, for refactoring tasks, structural verification scripts. These artifacts are executed only after task completion to validate additional scenarios and confirm that the requested refactoring has actually been performed.

Bug-fix tasks contain at least one public test that fails in the initial version and passes after a correct implementation. Feature tasks start with all existing tests passing, while newly added hidden tests validate the requested functionality. Finally, refactoring tasks require all behavioral tests to pass both before and after the modification. In these cases, the scripts located in the \texttt{evaluation} directory additionally verify structural properties of the resulting code, such as the existence of a newly extracted function, the movement of a function between files, or the complete removal of an obsolete public API.\\

\graybox{
{\em Availability:} \arkbench\ is publicly available at \url{https://github.com/mtov/arkbench}.}.

\subsection{Evaluation Results}
\label{sec:results}

Table~\ref{tab:ark-results} summarizes the results obtained by \ark\ on \arkbench. For each task, we report whether the task was completed successfully, together with the total number of tokens consumed, the execution time, and the number of tool invocations performed during the agentic loop. Overall, \ark\ successfully solved 8 out of the 10 benchmark tasks (80\%). All four feature implementation tasks were completed successfully, as well as two of the three bug-fix tasks and two of the three refactoring tasks. The average execution time was 23.5 seconds per task, with most tasks completing in less than 30 seconds. On average, the agent consumed 3,342 tokens and performed 7.1 tool invocations per task, indicating that the benchmark can be executed efficiently while still requiring multiple interactions with the code base. It is also important to note that the results reported in Table~\ref{tab:ark-results} were obtained after a single execution of \ark. Thus, since LLMs are non-deterministic, they may vary across runs.

\begin{table}[ht]
\centering
\small
\caption{Results of \ark\ on \arkbench.}
\label{tab:ark-results}
\begin{tabular}{lcccc}
\toprule
\textbf{Task} &
\textbf{Result} &
\textbf{Tokens} &
\textbf{Time (s)} &
\textbf{\# Tool Calls} \\
\midrule

\texttt{bugfix-01}   & Success & 2,430 & 18.04 & 6 \\
\texttt{bugfix-02}   & Fail    & 4,176 & 63.94 & 8 \\
\texttt{bugfix-03}   & Success & 2,315 & 14.84 & 6 \\

\addlinespace

\texttt{refactor-01} & Success & 4,024 & 24.14 & 8 \\
\texttt{refactor-02} & Success & 6,584 & 18.99 & 10 \\
\texttt{refactor-03} & Fail    & 4,461 & 23.16 & 9 \\

\addlinespace

\texttt{feature-01}  & Success & 2,422 & 26.73 & 6 \\
\texttt{feature-02}  & Success & 2,207 & 14.19 & 6 \\
\texttt{feature-03}  & Success & 2,441 & 15.75 & 6 \\
\texttt{feature-04}  & Success & 2,360 & 14.94 & 6 \\

\midrule
\textbf{Average} &
\textbf{80\%} &
\textbf{3,342} &
\textbf{23.47} &
\textbf{7.1} \\
\bottomrule
\end{tabular}
\end{table}

The two unsuccessful tasks illustrate different challenges:

\begin{itemize}

\item In \texttt{bugfix-02}, the agent produced only a partial repair. It correctly handled the invalid-input cases covered by the visible tests, allowing all public tests to pass. However, when the customer object lacked an {\tt id} field, the patch still raised a {\tt KeyError} instead of the expected {\tt ValueError}. Thus, the issue was the use of a wrong exception type. 

\item In \texttt{refactor-03}, the patch was applied successfully, but the rename was not propagated throughout the codebase. Although the agent updated the function definition and the visible unit test, it failed to update all imports and call sites, leaving one module dependent on the old name. Consequently, the public tests passed, but the hidden evaluation failed during test collection with an import error. This result illustrates a common limitation of LLM-based code agents in refactoring tasks: they often perform the primary transformation correctly while overlooking secondary dependencies, producing patches that are incomplete.

\end{itemize}

\noindent{\bf Tool Usage Analysis: }
To better understand how \ark\ solves programming tasks, we analyzed the sequence of tool invocations produced during the execution of the benchmark tasks. Although the exact sequences varied across tasks, we observed a recurring workflow that closely mirrors the architecture presented in Section~\ref{sec:architecture}. In particular, 6 out of the 10 benchmark executions (60\%) followed the pattern shown in Figure~\ref{fig:tool-pattern}, starting with repository exploration (\texttt{list\_files}), followed by context acquisition (\texttt{read\_file}), solution generation (\texttt{finish}), patch application (\texttt{apply\_patch}), and validation through automated tests (\texttt{run\_tests}). The remaining executions followed variations of this workflow, mainly involving additional repository exploration or file inspection steps.

\begin{figure}[ht]
\centering
\fbox{
\parbox{0.85\linewidth}{
\centering
\texttt{list\_files}
$\rightarrow$
\texttt{read\_file}
$\rightarrow$
\texttt{finish}
$\rightarrow$
\texttt{apply\_patch}
$\rightarrow$
\texttt{run\_tests}
}
}
\caption{Representative tool invocation pattern.}
\label{fig:tool-pattern}
\end{figure}

This pattern provides empirical evidence that the main architectural components of \ark\ were consistently exercised during benchmark execution. More importantly, it highlights the central role of validation in the agentic workflow. Rather than directly committing generated code, \ark\ first materializes the proposed solution as a patch and then executes the test suite to verify its correctness. This generate--apply--validate cycle constitutes the core execution mechanism of the agent and was observed across all benchmark tasks.\\[-0.2cm]

\noindent{\bf Cost Analysis:} 
The entire evaluation of \ark\ on \arkbench\ using the model {\tt gpt-5.4-mini}  cost less than US\$0.50 in API usage. This result highlights the efficiency of \ark's minimalist architecture: despite solving 8 out of 10 maintenance tasks, the agent required relatively few iterations, tool invocations, and tokens. \\[-0.2cm]

\noindent{\bf Threats to Validity:}
The results reported in this section should be interpreted in light of some limitations. First, all experiments were conducted using a single LLM, and the observed behavior may not generalize to other models. Second, the evaluation relied on \arkbench, a lightweight benchmark comprising only ten tasks, which may not capture the full complexity of real-world software engineering scenarios. Third, each task was executed only once, and therefore the results may vary across runs due to the non-deterministic nature of LLMs. Finally, both \ark\ and \arkbench\ focus exclusively on Python projects, limiting the generalizability of the findings to other programming languages.\\

\graybox{
{\em Summary:} We evaluated \ark\ using \arkbench, a lightweight benchmark comprising ten representative software maintenance and evolution tasks. \ark\ successfully solved 8 of the 10 tasks while maintaining low execution time, modest token consumption, and an API cost below US\$0.50.}

\section{Comparison with Other Coding Agents}
\label{sec:comparison}

Recently, Rombaut proposed an architectural taxonomy for coding agents based on an analysis of 13 open-source systems~\cite{Rombaut2026}. The taxonomy organizes code agent architectures along twelve dimensions grouped into three layers. The first layer, \textit{Control Architecture}, describes how an agent structures its execution flow and makes decisions. The second layer, \textit{Tool and Environment Interface}, characterizes the agent's interaction with the development environment, including the available tools, code editing mechanisms, and command execution facilities. Finally, the \textit{Resource Management} layer describes how the agent manages internal resources such as state, context, memory, and LLMs.

In this section, we use this taxonomy to compare the architecture of \ark\ with that of other coding agents. Rather than considering all the systems analyzed by Rombaut, however, we focus on two representative agents. The first is \agent{Codex CLI}, developed by OpenAI, which represents the state of the practice in commercial coding agents. The second is \agent{OpenCode}, one of the most popular open-source coding agents, representing projects developed and maintained by a broader community. The goal of this comparison is to position \ark\ within the architectural landscape of modern coding agents, highlighting its main design decisions and the simplifications introduced to make its architecture easier to understand and better suited for research and education.

\subsection{Control Architecture}

In Rombaut's taxonomy, the \textit{Control Architecture} layer describes how an agent organizes its execution flow, that is, how it decides which action to perform next and how it controls its interaction with the LLM. This layer comprises three dimensions: the type of control loop, the entity responsible for driving the loop, and the way the loop is implemented. Table~\ref{tab:control-comparison} compares \ark\ with \agent{Codex CLI} and \agent{OpenCode} along these three dimensions.\footnote{The \agent{Codex CLI} and \agent{OpenCode} columns in this and in the following tables are reproduced from Rombaut~\cite{Rombaut2026}.}
As shown, all three agents share the same control architecture. They implement a sequential ReAct-based loop in which the LLM itself decides, at each iteration, which tool should be executed next. Furthermore, all three rely on an imperative implementation based on a \texttt{while} loop to coordinate the successive interactions between the LLM and the available tools. These results show that, although \ark\ is substantially smaller and simpler than the other two agents, it adopts the same control architecture.

\begin{table*}[ht]
\centering
\small
\caption{Control architecture.}
\label{tab:control-comparison}
\begin{tabularx}{\linewidth}{lXXX}
\toprule
 & \textbf{Codex CLI} & \textbf{OpenCode} & \textbf{Ark}  \\
\midrule
Control loop & ReAct Loop & ReAct Loop & ReAct Loop \\
Loop driver & LLM-driven & LLM-driven & LLM-driven \\
Control flow implementation & Imperative \texttt{while} loop & Imperative \texttt{while} loop & Imperative \texttt{while} loop \\
\bottomrule
\end{tabularx}
\end{table*}

\subsection{Tool and Environment Interface}
\label{sec:tool-comparison}

This layer describes how the agent interacts with the local development environment. It encompasses the set of tools available to the LLM, the mechanism used to apply code modifications, how tools are exposed to the model, the mechanisms used for context retrieval, and the degree of isolation of the execution environment. Table~\ref{tab:tools-comparison} compares these characteristics across \agent{Codex CLI}, \agent{OpenCode}, and \ark.

\begin{table*}[ht]
\caption{Tool and environment interface.}
\label{tab:tools-comparison}
\centering
\small
\begin{tabularx}{\linewidth}{lXXX}
\toprule
 & \textbf{Codex CLI} & \textbf{OpenCode} & \textbf{Ark} \\
\midrule
Tool set size & 20+ & 18+ & 5 \\

Edit and patch format & Apply patch & String replacement and apply patch & Apply patch \\

Tool discovery & Per-turn rebuild & Dynamic & Static \\

Context retrieval & Keyword/regex search & Keyword/regex search & Keyword search \\
Execution isolation & Platform sandboxing & Local shell & Local shell \\
\bottomrule
\end{tabularx}
\end{table*}

As shown in the table, the three agents differ substantially in their tool and environment interfaces. First, both \agent{Codex CLI} and \agent{OpenCode} expose more than 18 tools to the LLM. By contrast, \ark\ provides only five tools, as described in Section~\ref{sec:tools}. This design choice reduces implementation complexity and makes the architecture more suitable for educational purposes. With respect to code editing, \agent{Codex CLI} and \ark\ both adopt the {\it apply patch} mechanism based on unified diffs, whereas \agent{OpenCode} supports both direct text replacement ({\it string replacement}) and {\it apply patch}. As discussed in Section~\ref{sec:tools}, the {\it apply patch} approach facilitates patch validation before changes are applied to the repository.

Another important difference lies in tool discovery. \agent{Codex CLI} reconstructs the set of available tools dynamically at each iteration, allowing new tools to be incorporated while the agent is running. Similarly, \agent{OpenCode} supports dynamic tool discovery through plugins and MCP servers. In contrast, \ark\ relies on a static set of tools defined when the agent is initialized, thereby simplifying the interaction protocol between the agent and the LLM. The agents also differ in their context retrieval mechanisms. Both \agent{Codex CLI} and \agent{OpenCode} support keyword and regular-expression searches, providing greater flexibility for locating relevant code. In contrast, \ark\ implements only keyword-based search, which proved sufficient for the evaluation scenarios considered in this work. Future studies may, however, demonstrate the value of more powerful search capabilities. Finally, \agent{OpenCode} and \ark\ execute commands directly through the local operating system shell, whereas \agent{Codex CLI} relies on platform-provided sandboxing mechanisms to reduce the risks associated with executing harmful commands. This difference reflects \ark's emphasis on architectural simplicity, whereas production-oriented agents incorporate more sophisticated isolation mechanisms.

\subsection{Resource Management}
\label{sec:resource-comparison}

This layer describes how an agent manages its resources during and across executions. In particular, it considers the representation of the agent's internal state, the strategies used to reduce the size of the context sent to the LLM, the use of specialized language models, and the availability of persistent memory mechanisms. Table~\ref{tab:resources} compares these characteristics across \agent{Codex CLI}, \agent{OpenCode}, and \ark.

\begin{table*}[ht]
\caption{Resource management.}
\label{tab:resources}
\centering
\small
\begin{tabularx}{\linewidth}{lXXX}
\toprule
 & \textbf{Codex CLI} & \textbf{OpenCode} & \textbf{Ark} \\
\midrule
State management & Flat list, preserved & Typed event log & Destructive \\
Context compaction & LLM-based summarization & LLM-based summarization & Size truncation \\
Multi-model routing & Safety-focused & Role-based & Single model (no routing) \\
Persistent memory & Background extraction pipeline & Full session persisted in SQLite & None \\
\bottomrule
\end{tabularx}
\end{table*}

As shown in the table, the three agents differ substantially in how they manage internal resources. First, with respect to state management, \agent{Codex CLI} maintains a linear history of messages throughout the session, whereas \agent{OpenCode} uses a typed event log, in which events represent actions, observations, and state changes. By contrast, \ark\ adopts a simpler representation. As described in Section~\ref{sec:context}, its state consists only of the recent interaction history, with older entries discarded as needed.

These differences are also reflected in the agents' context compaction strategies. Both \agent{Codex CLI} and \agent{OpenCode} rely on the LLM itself to summarize the execution history when the context window approaches its limit. In contrast, \ark\ simply discards the oldest entries once the interaction history exceeds a predefined maximum size. Although less sophisticated, this approach is simple, predictable, and well suited to a minimal coding agent. Another important difference concerns the use of multiple language models. \agent{Codex CLI} employs a specialized model to assess the risks associated with certain actions. Similarly, \agent{OpenCode} supports model routing, allowing different models to perform specialized roles during execution. By contrast, \ark\ relies on a single model throughout the entire execution. Finally, the three agents differ in their support for persistent memory. \agent{Codex CLI} stores relevant information in the background for reuse in future sessions. \agent{OpenCode} persists the entire session state in an SQLite database, enabling previous executions to be resumed. In contrast, \ark\ does not retain any information across executions: its entire memory is discarded when the task completes.\\

\graybox{
{\em Summary:} The comparison shows that \ark\ preserves the core architectural elements found in modern coding agents, particularly the ReAct-based control architecture driven by the LLM itself. The main differences lie in the Tool and Environment Interface and Resource Management layers, where \ark\ adopts simpler design choices, including a small predefined tool set, keyword-based context retrieval, the absence of persistent memory, and the use of a single model. These simplifications result in a smaller implementation that is easier to understand, modify, and extend.}

\section{Related Work}
\label{sec:related-work}

Research on coding agents has expanded rapidly in recent years, spanning topics such as software maintenance and repair~\cite{wang2025aegis,bouzenia2025repairagent}, software design analysis~\cite{batole2025localizeagent}, automated testing~\cite{fan2024marg}, and the empirical study of agent behavior~\cite{bouzenia2025understanding,hassan2026se30}. In this section, we position \ark\ with respect to the most closely related work. We first discuss studies that investigate the architecture of coding agents, including architectural taxonomies and reference models. We then review research-oriented coding agents proposed in the literature and conclude with recent studies that analyze the practical use and configuration of coding agents in software engineering.

\subsection{Software Architecture \& Coding Agents}

The work most closely related to ours is that of Rombaut~\cite{Rombaut2026}, who proposed an architectural taxonomy for coding agents based on an analysis of 13 open-source systems. The taxonomy characterizes these agents along twelve dimensions organized into three layers, providing a comprehensive framework for describing the main architectural design decisions of modern coding agents. In this paper, we adopt this taxonomy as the basis for the comparative analysis presented in Section~\ref{sec:comparison}. However, whereas Rombaut focuses on comparing existing architectures, we present \ark, an open-source implementation that embodies the essential architecture of a coding agent and serves as a reference implementation for research and education.

Reference architectures are high-level architectural models that capture the common components, responsibilities, and relationships found across a family of systems. Rather than describing a specific implementation, they provide a shared blueprint for understanding, comparing, evaluating, and designing systems within a domain. For example, Liu and David proposed a reference architecture for reinforcement learning (RL) frameworks based on an empirical analysis of 18 existing frameworks, identifying recurring components such as {\em Agent}, {\em Environment}, {\em Framework Orchestrator}, {\em Data Persistence}, and {\em Monitoring \& Visualization} and organizing them into a unified architectural model~\cite{liu2026reference}.   In contrast, \ark\ is not a reference architecture but a concrete implementation of a coding agent. While a reference architecture seeks to capture the common architectural structure of a domain, \ark\ materializes and validates a particular architectural design through an executable research prototype. 

\subsection{Research-Oriented Coding Agents}

Table~\ref{tab:related-agents} summarizes the main differences between \ark\ and selected research-oriented coding systems. These systems and their key differences from \ark\ are discussed throughout the remainder of this subsection.

\begin{table*}[ht]
\centering
\small
\caption{Comparison of research-oriented coding systems with \ark}
\label{tab:related-agents}
\begin{tabular}{p{3cm} p{12cm}}
\toprule
\textbf{Agent} & \textbf{Key difference from \ark} \\
\midrule

SWE-agent &
Focuses on task-solving performance, whereas \ark\ emphasizes architectural clarity, simplicity, and completeness. \\

mini-swe-agent &
Provides an extremely minimal implementation, while \ark\ includes explicit memory, history management, and patch validation. \\

AutoCodeRover &
Uses specialized techniques such as AST-based search and fault localization, whereas \ark\ adopts a minimal, general-purpose architecture. \\

OpenHands &
Provides a comprehensive platform with sandboxing, web browsing, multi-agent support, and evaluation infrastructure, whereas \ark\ is a compact implementation. \\

Agentless &
Uses a fixed localization--generation--validation pipeline, whereas \ark\ follows an iterative agentic loop with dynamic tool use. \\

\bottomrule
\end{tabular}
\end{table*}
SWE-agent~\cite{yang2024sweagent} was one of the first systems to demonstrate that coding agents can autonomously solve software maintenance tasks in real-world repositories, particularly issues involving bug fixes and small code modifications. The agent operates through a ReAct loop in which the LLM selects actions over an Agent-Computer Interface (ACI), such as navigating the repository, searching for information, editing files, and executing commands, while using the resulting observations to guide subsequent decisions. Its success influenced many later coding agents, helping establish the architecture based on an agentic loop and tool execution. \ark\ follows the same architectural model but places emphasis on architectural clarity and implementation simplicity. For example, 37\% of its commits were devoted to refactoring aimed at making both the codebase and the overall design easier to understand. More recently, the authors of SWE-agent released \texttt{mini-swe-agent}, a reimplementation of approximately 100 lines of code intended to illustrate the core principles of the original system. Although remarkably elegant, this implementation abstracts away several components that are implemented in \ark, including memory and history management, patch validation and application. In other words, \ark\ was carefully designed to strike a balance between simplicity, architectural clarity, and completeness.

AutoCodeRover~\cite{zhang2025autocoderover} is a coding agent designed to resolve GitHub issues. Its distinguishing feature is the combination of LLMs with classical software engineering techniques, such as AST-based structural search and spectrum-based fault localization, to identify the relevant code before generating a {patch. By contrast, \ark\ prioritizes a minimal, general-purpose architecture rather than incorporating specialized techniques for code localization and repair.

OpenHands (formerly known as OpenDevin) is an open-source platform for research and development on software agents, designed to serve both as an execution environment and as an experimental infrastructure for agent-based systems~\cite{openhands2025}. The platform provides a comprehensive set of capabilities, including sandboxed command execution, code editing, web browsing, multi-agent coordination, and an evaluation framework based on benchmarks such as SWE-Bench and WebArena. In contrast, \ark\ follows a deliberately minimalist philosophy. Rather than providing a comprehensive platform for developing and evaluating coding agents, its goal is to offer an educational and easy-to-understand reference implementation of a coding agent's architecture.

Agentless~\cite{xia2025agentless} investigates whether coding agents based on iterative loops and dynamic tool use are truly necessary for solving software engineering tasks. Rather than adopting an agentic architecture, the system employs a fixed execution pipeline consisting of code localization, patch generation, and validation. Although this approach achieves competitive results on benchmark tasks, agent-based architectures have emerged as the dominant paradigm for programming assistants, demonstrating generality and flexibility to support a wide range of tasks.

Recent work has begun to investigate how software engineering agents behave internally by analyzing their thought--action--result trajectories~\cite{bouzenia2025understanding}. Such studies provide insights into reasoning patterns, tool usage, and common failure modes. Following the same motivation, \ark\ records detailed execution traces, enabling future investigations of coding-agent behavior and decision-making processes.
Other work argues that software engineering agents should move beyond purely reactive interactions by incorporating structured reasoning, persistent state, and execution-grounded feedback mechanisms~\cite{chen2026-boatse}. Research prototypes such as \ark\ can serve as experimental platforms for investigating and refining these principles in practice, enabling controlled studies of memory, reasoning, and validation mechanisms in coding agents.

\subsection{Other Works}

The OpenAI Agents SDK is an open-source framework for building agent-based applications~\cite{openai_agents_sdk}. It provides high-level abstractions for defining agents, tools, task handoffs, guardrails, and tracing. Rather than manually implementing the interaction loop between the model and external tools, developers can define specialized agents and delegate execution coordination to the framework. The SDK also supports multi-agent systems, asynchronous execution, MCP integration, context management, and observability. While the coding agents discussed previously in this section are specialized for software engineering tasks, the OpenAI Agents SDK is a general-purpose framework for developing agents across different application domains. 

Two recent studies from our research group investigate coding agents from complementary perspectives. Santos et al.~\cite{agent2026-claude-code} analyzed 328 configuration files (\texttt{CLAUDE.md}) from public repositories, identifying the main software engineering concerns used to guide the behavior of coding agents, as well as the most common configuration patterns. Almeida et al.~\cite{agent2026-copilot} evaluated GitHub Copilot Agent Mode for the automated migration of the SQLAlchemy library across ten real-world applications. Their results showed that, although the agent achieved high migration coverage, it still struggled to preserve the functional behavior of the migrated systems. These findings suggest several directions for future work with \ark. For example, it would be interesting to investigate how project-specific configuration files, such as {\tt AGENTS.md} and {\tt CLAUDE.md}, affect the behavior and performance of coding agents. Another promising direction is to specialize \ark\ for specific software engineering tasks, such as automated code migration.

Our work on \ark\ builds on earlier efforts to rethink how developers interact with language models during software development. In previous work, we proposed NoCodeGPT, a customized interface that enabled users to build web applications through prompts without directly manipulating source code~\cite{nocodegpt}. At the time, coding agents had not yet emerged, but we had already observed that general-purpose chat interfaces were not well suited for software construction. 

\section{Conclusion}
\label{sec:conclusion}

Coding agents are rapidly becoming a fundamental interface for software development. Despite their growing importance, however, their architecture has received relatively little attention from the software engineering community. This paper addressed this gap by presenting a systematic description of the main architectural components of coding agents, including the agentic loop, system prompt, tools, memory, context management, error handling, security mechanisms, tracing, and LLM integration. Our goal was to explain the architectural principles underlying modern coding agents and how these components cooperate to solve software engineering tasks.

To make these concepts concrete, we presented \ark, a minimal open-source coding agent designed for research and education. We also introduced \arkbench, a lightweight benchmark comprising ten representative software maintenance and evolution tasks. The experimental results showed that \ark\ successfully solved 8 of the 10 benchmark tasks while maintaining low execution time, and modest token consumption. Finally, by comparing \ark\ with state-of-the-art coding agent, we showed that a minimal implementation can preserve the core architectural principles of modern coding agents while remaining smaller and easier to understand.

Our experience implementing \ark\ also provided several practical insights into the design of coding agents:

\begin{enumerate}

\item Although the agentic loop is the architectural core of a coding agent, most implementation complexity lies in supporting components such as context management, memory, security, and error handling.
    
\item The system prompt proved to be as important as the implementation itself, since small prompt refinements often had a substantial impact on the agent's behavior.
    
\item Patch validation emerged as an essential mechanism for handling imperfect LLM outputs and ensuring reliable code modifications.
    
\item Context management directly affects both cost and performance, requiring careful trade-offs between providing sufficient information and controlling token consumption.
    
\item A simple architecture can remain effective while offering advantages in terms of explainability, maintainability, and suitability for research and education.
\end{enumerate}

Several directions remain for future work. First, \ark\ can be extended with capabilities commonly found in production-oriented agents, such as persistent memory, richer tool sets, more sophisticated context management, and multi-model routing. Second, \arkbench\ can be expanded with additional maintenance scenarios and larger projects to support more comprehensive evaluations. Finally, we hope that both \ark\ and \arkbench\ will serve as useful platforms for teaching, experimentation, and future research on the architecture and engineering of coding agents.

\subsection*{Acknowledgments}

This research is supported by grants from FAPEMIG and CNPq.

\bibliographystyle{plain}
\bibliography{references}

\end{document}